\documentclass[prb,twocolumn,showpacs,aps]{revtex4}
\usepackage {graphicx}
\pdfoutput=1
\begin{document}
\title{Enhanced magnetism, memory and aging in Gold-Iron oxide nanoparticle composites}

\author{S. Banerjee$^a$\footnote{Email:sangam.banerjee@saha.ac.in}, S. O. Raja$^b$, M. Sardar$^c$, N. Gayathri$^d$, B. Ghosh$^a$, A. Dasgupta$^b$}

\address{$^a$ Surface Physics Division,  Saha Insitute of Nuclear Physics, 1/AF Bidhannagar, Kolkata 700 064, India\\ $^b$ Department of Biochemistry, University of Calcutta, 35 B. C. Road, Kolkata-700019,\\ $^c$ Material Science Division, Indira Gandhi Center for Atomic Research, Kalpakkam 603 102, India,\\ $^d$ Material Science Section, Variable Energy Cyclotron Center, 1/AF Bidhannagar, Kolkata 700 064, India}

\begin{abstract} 
In this report we present systematic magnetic studies of pure iron
oxide nanoparticles and gold-iron oxide nanocomposite with increasing
Au particle size/content. For the magnetic studies of these samples we
have measured: (1) zero field cooled (ZFC) and field cooled (FC)
magnetization, (2) ac susceptibility, (3) magnetization vs field at
various temperatures, (4) thermoremanant magnetization relaxation
(TRM) and zero field cooled magnetization relaxation (ZFCM) at fixed
temperature for various wait times $t_w$ for studying the aging effect, (5) magnetization memory effect and (6) exchange bias as a function of cooling field. The detailed magnetic measurement
analysis indicates that the pure Fe$_3$O$_4$ nanoparticles sample
behaves like a superparamagnet and on incorporation of gold (Au)
nanoparticles the nanocomposite system slowly evolves from
superparamagnetic to superspin glass state. The memory and aging
effect enhances with the increase of the Au nanoparticle size/content.
The most important observation in this study is the enhancement of
magnetization with the incorporation of Au nanoparticles. The
enhancement increases with the increase in the Au content in the
nanocomposite. We have explained the cause of this enhancement of
magnetization as due to large orbital magnetic moment formation at the Au/magnetic particle interface.

\end{abstract}

\maketitle
\section{Introduction}
Understanding the static and dynamic behaviour of an ensemble of magnetic nanoparticles (NPs) has become a subject of intense research interest in 
recent years\cite{bedanta,dormann1,dormann2}. Their rich contribution
to fundamental physics and their importance in technological application has become well established now \cite{freitas}. A variety of competing interparticle interactions among the magnetic NPs can give rise to unusual experimental phenomena. It is well known that
 when a bulk magnetic multi-domain specimen is reduced below a critical size, the particles becomes magnetically single domain and acquires a giant spin \cite{neel}. The magnetic behavior of these ensemble of NPs critically depends on the competition between the magnetic anisotropy energy of an individual nanoparticle and the magnetic dipole-dipole interaction between the particles. If the former is higher then the dynamic behaviour follows the N\'{e}el-Brown model \cite{neel,brown} and the system is termed as superparamagnetic (SPM)\cite{cullity} exhibiting magnetic viscosity due to N\'{e}el relaxation \cite{neel}. If the dipole-dipole interaction is of the order of the particle anisotropy energy then the system can go to a magnetically fustrated state leading to a system termed as superspin glass (SSG)\cite{kleemann,sun,sasaki,suzuki}. Both the systems mentioned above have a characteristic temperature; T$_B$ or blocking temperature in the case of SPM NPs and T$_f$ or spin freezing temperature in the case of SSG NPs. Below this characteristic temperature, the magnetic moments are frozen and the system exhibits non-equilibrium properties like memory effect and magnetic hysteresis \cite{kleemann,sun,sasaki,suzuki,tsoi}. Additionally, the SSG system exhibits \textquoteleft aging effect \textquoteright \cite{sasaki,suzuki,sahoo} which is absent in the case of SPM NPs. Unusually slow dynamical behaviours like memory are explained within mainly two paradigms, (1) spin glass state arising from frustrated interparticle interaction and disorder \cite{sasaki,suzuki,chen,frus,giri1}, (2) freezing of superparamagnets with unavoidable polydispersity \cite{duttagupta1,duttagupta2}. It is also known that metallic NPs also exhibits novel electronic, optical and magnetic properties. It has been reported that 1.7 nm gold NPs surrounded by thiol shows ferromagnetic hysteresis at room temperature, which has been attributed to orbital magnetism \cite{authiol,hernando}. Recently ferromagnetism in graphite \cite{graphite}, non-magnetic oxides and borides \cite{sangamapl,oxides,borides} have been reported and the ferromagnetic hysteresis observed in these systems have also been attributed to orbital magnetism \cite{sangamapl,simu} occuring due to nanosize defects/structures.  The ensemble of magnetic nanoparticle system becomes more rich in physics when metallic or semiconducting nanoparticles are incorporated. More recent trend in magnetic nanoparticles research, is to synthesize assemblies of different materials like (1) core-shell nanoparticles where magnetic (non-magnetic) particles are encapsulated by non-magnetic (magnetic) layers \cite{srikanth1}, (2) multicomponent nanoparticles with mixtures of magnetic and nonmagnetic nanoparticles in physical contact (nanocomposites)\cite{du,srikanth2}.

Nobel metal gold (Au) composites with thiol capping \cite{authiol} or Fe$_3$O$_4$ \cite{srikanth1,srikanth2} have shown unusual magnetic
properties when prepared in the nanoforms. For example, in the case of thiol capped Au, the composite becomes magnetic \cite{authiol}. When Fe$_3$O$_4$ is capped with Au the magnetization value of the composite decreases,  but the blocking temperature does not show any consistent behaviour \cite{srikanth1,wang}.
For a review of the magnetic nature, single domain
limit of Fe$_3$O$_4$ nanoparticles please see\cite{caruntu}.
 In this paper we have carried out a systematic study to understand the role of gold in the magnetic behavour of Fe$_3$O$_4$ NPs. In any of the works reported earlier, no study has been done as a function of increasing Au particle size/content. In this paper we report on the synthesis, structure and magnetic characterization of  nanocomposites of Fe$_3$O$_4$ NPs and Au NPs with increasing particle size of Au. The main findings of this present study are (1) Blocking temperature can be widely varied by changing the size and amount of Au, (2) Magnetization shows deviation from superparamagnetic scaling for $T\gtrsim 300K$ and this deviation
increases as the Au content increases, (3) The frequency dependent magnetic susceptibility shows activated (N\'{e}el-Brown-Arrhenius-Vogel
-Fulcher) relaxation for pure Fe$_3$O$_4$ and sample with low Au content, but evolves to simple power law type relaxation for sample with high Au content, (4) Memory effect is seen in the field cooled magnetization measurements, and this memory effect increases with Au content and (5) A large increase in magnetic moment with increase of Au particle size/content which is rather unusual contradicting the earlier reports. In this paper we shall try to address all the above
points and explain our observation and particularly try to explain the most important observation of this present investigation ,
 i.e the enhancement of magnetisation upon incorporation of Au. 

\section{Experimental details}
Fe$_3$O$_4$ NPs were initially prepared by co-precipitation method. 4 gm ferric chloride and 2 gm ferrous chloride (2:1, w/w ratio) were dissolved in 2 M HCl and co-precipitated by 100 ml 1.5 M NaOH solution upon constant stirring for 30 minutes at room temperature. The prepared colloidal solution was centrifuged to collect the supernatant (suspendend) solution to obtain particles with a narrow size distribution. The supernatant solution was pelleted down by a strong magnet and washed four times by ultra pure water. Finally 20 ml Citrate buffer (1.6 gm Citric acid and 0.8 gm tri-sodium citrate) was added to collect the stabilized ferrofluid in solution at a pH around 6.3. This solution was used as a base in the subsequent prepartion of the nanocomposite samples. The solution was lyophilized to obtain the pure Fe$_3$O$_4$ sample which will be referred to subsequently as Sample A. The following procedure was adopted to prepare the Au:Fe$_3$O$_4$  nanocomposite samples: 300$\mu$L of the synthesized colloidal iron oxide nanoparticle (~0.1M) suspension was added to 25ml ultra pure boiling water under vigorous stirring condition. Then 350$\mu$L of 20mM HAuCl$_4$ is added and finally 300$\mu$L of 100mM Tri-sodium citrate was added. The whole solution was kept boiling and stirred for 15 minutes till the color of the solution turned from black to red. This red solution was further centrifuged to obtain two samples with different Au NP's sizes keeping the Fe$_3$O$_4$ particle size same. The supernatant and the pellet solution were lyophilized to obtain the dry Low-Au sample (Sample B) and the High-Au sample (Sample C) respectively. The particle size of the Au NPs in the Low-Au sample was small and that of the High-Au were big comparitively. The samples were characterized using high resolution transmission electron microscopy (TEM). The magnetic property of all the samples were measured using MPMS-7 (Quantum Design). We have measured (1) zero field cooled (ZFC) and field cooled (FC) magnetization at 100 Oe, (2) ac susceptibility using 3 Oe field at 3.3Hz, 33Hz, 90Hz and 333Hz and 1kHz, (3) magnetization vs field upto 5 Tesla at 5K, 125K, 200K and 300K, (4) thermo remanent magnetization relaxation (TRM) and zero field cooled magnetization relaxation (ZFCM) behaviour at 50 Oe at 80K for various wait times $t_w$ = 300 secs, 3000 secs and 3 hours, (5) magnetization memory effect with the following protocol - the magnetization was measured during the field cooling at 100 Oe and at certain temperatures the field was switched off for 2 hours. After the wait time, the field was again switched on and the magnetization measurement was carried out subsequently during cooling till the next stop temperature was reached. This was carried out down to 5K. During warming, the magnetization was measured without any break and (6) exchange bias as a function of cooling field of 100 Oe, 500 Oe, 0.2T and 1T at 10K.

\section{Results and Discussion}
In fig~1 we show the TEM micrograph of the (a) Low-Au and (b-d) High-Au samples. The particles appearing with lower contrast are Fe$_3$O$_4$ particles and those with high contrast (dark) are the Au particles. In both the samples the Fe$_3$O$_4$ particles are typically 3-4 nm in size. The Au particles in the Low-Au sample are nearly monodispered with particle size $\sim$5-6 nm whereas in the high-Au sample, they are polydispersed with the particle sizes ranging from $\sim$ 7-10 nm. In the low magnification micrographs fig.1(c-d) of Sample C, we also observed very large particles ($\sim$ 200 nm in size) having core-shell structures with Fe$_3$O$_4$ at the core and Au as the outer shell.

\begin{figure}
\includegraphics*[width=9.5cm]{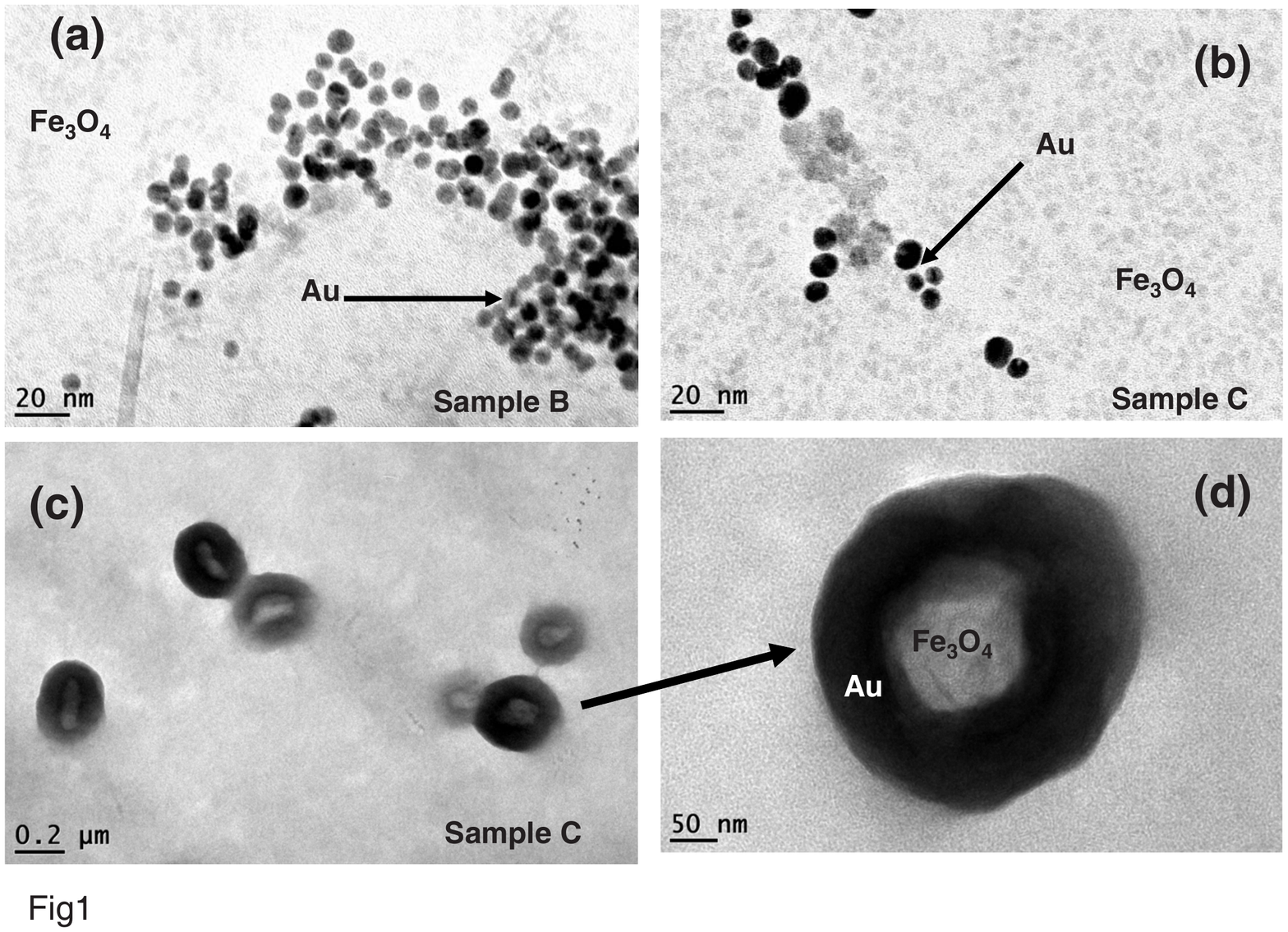}
\caption{(See online for better contrast of Fe$_3$O$_4$ particles) Transmission Electron Micrographs of the (a) Low-Au (Sample B) and (b-d) High-Au (Sample C) nanocomposites. Fe$_3$O$_4$ can be seen in the background as faint particles of size 3-4 nm. Au particles are darker and are marked by arrows. Core(Fe$_3$O$_4$)-shell(Au) structures seen in Sample C are shown in (c) and (d). }
\end{figure}

In fig~2 we show the field cooled (FC) and zero field cooled (ZFC)
magnetization variation with temperature for all the three samples.
The first and foremost observation is that there is an  increase in
the magnetization value with increasing Au content. The high-Au sample
shows a much larger enhancement of the magnetization compared to the
low-Au sample. In addition the blocking  temperature (the temperature
at which the ZFC curve peaks) shifts towards higher temperature as the
Au particle size increases (T$_B$ $\sim$ $35$ K, $80$ K and $180$K
for pure Fe$_3$O$_4$, Low-Au and High-Au samples respectively). The
other interesting feature is the systematic differences in the
temperature dependence of the FC magnetization of the samples below
the ZFC peak temperature. Sample A shows the typical SPM behaviour
(increasing M with decreasing T following the usual paramagnetic T
dependence). Sample B shows a less steeper temperature dependence and
Sample C shows almost a constant magnetization (no temperature
dependence) at lower temperatures typical of a (super) spin glass
\cite{sasaki,suzuki,sahoo,du,bitoh}. For a superspin glass it is known
that the FC magnetization either remains constant or decreases as a
function of decreasing temperature in contrast to the superparamagnet
where it increases \cite{sasaki,suzuki,sahoo,du,bitoh}.

\begin{figure}
\includegraphics*[width=9.5cm]{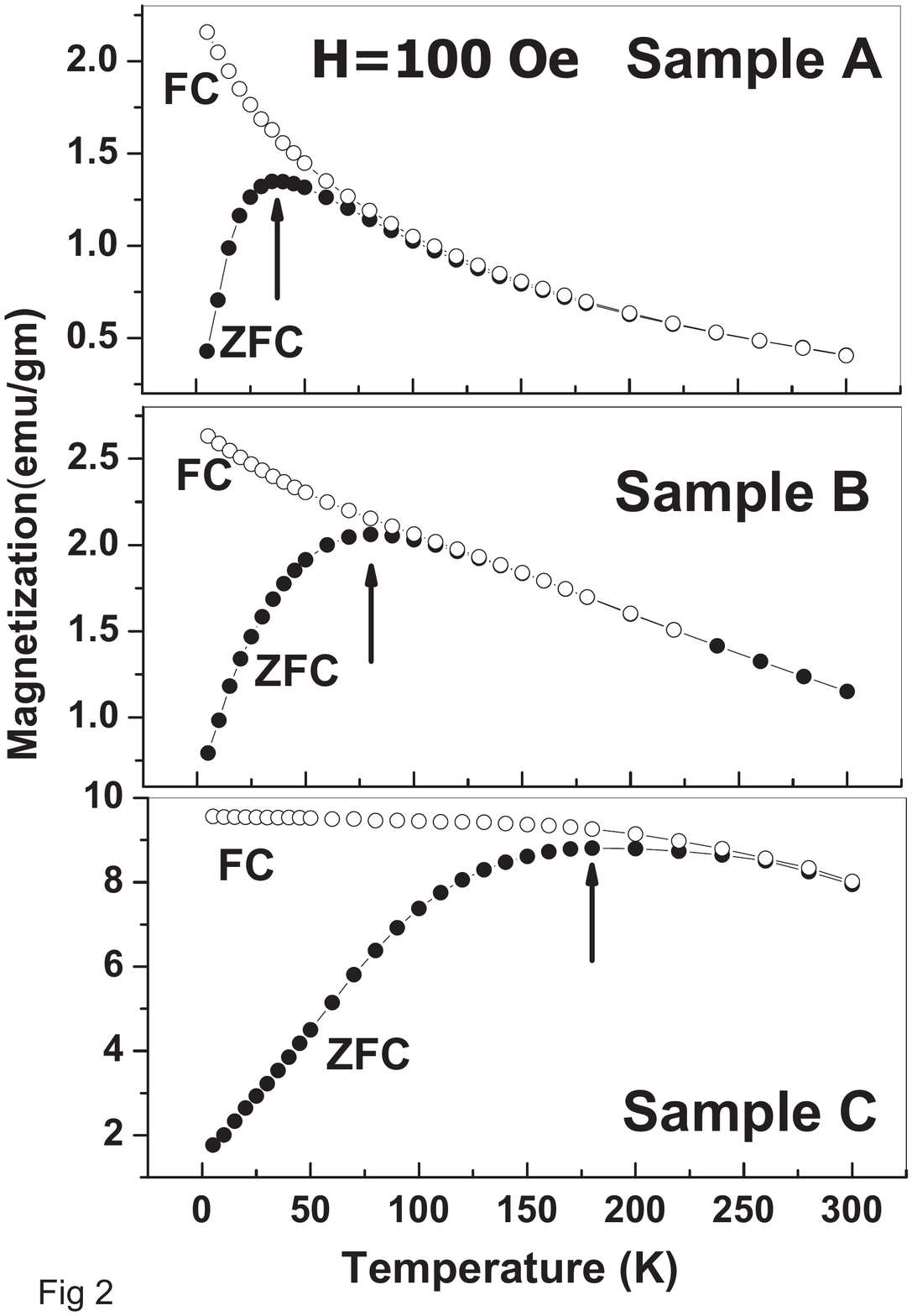}
\caption{Zero field cooled (ZFC) and Field Cooled (FC) magnetization curves for (a) Sample A - pure Fe$_3$O$_4$ sample, (b) Sample B - Low Au and (c) Sample C - High Au taken at a field of 100 Oe. The arrows show the peak position in the ZFC curves.}
\end{figure}

In fig~3 we have shown, magnetization versus H/T plot for three
different temperatures ($125$ K, $200$ K and $300$ K - we have
measured M vs H at 10K also, it shows hysteresis). While the
magnetization behaviour at the lower two temperatures scales with H/T
(falling on each other), the curve at $300$ K is consistently lower
which is an anamolous behaviour in these type of nanocomposite
systems. The deviation of magnetic scaling with H/T increases with the increase in Au content. This absence of scaling of the magnetization curves with H/T
seems to imply that all the three samples are not pure
superparamagnets above the peak temperature in ZFC as far as
magnetization is concerned. We also observe in fig~3, that there is a
slight enhancement in the saturation magnetization of the Low-Au
sample (sample B) compared to the pure sample (sample A), whereas the
saturation magnetization of the High-Au (Sample C) has enhanced
drastically. This large enhancement of magnetization observed for the
high-Au sample in fig~2(c) and fig~3 is quite unusual, and is the
central result of this investigation.

\begin{figure}
\includegraphics*[width=9.5cm]{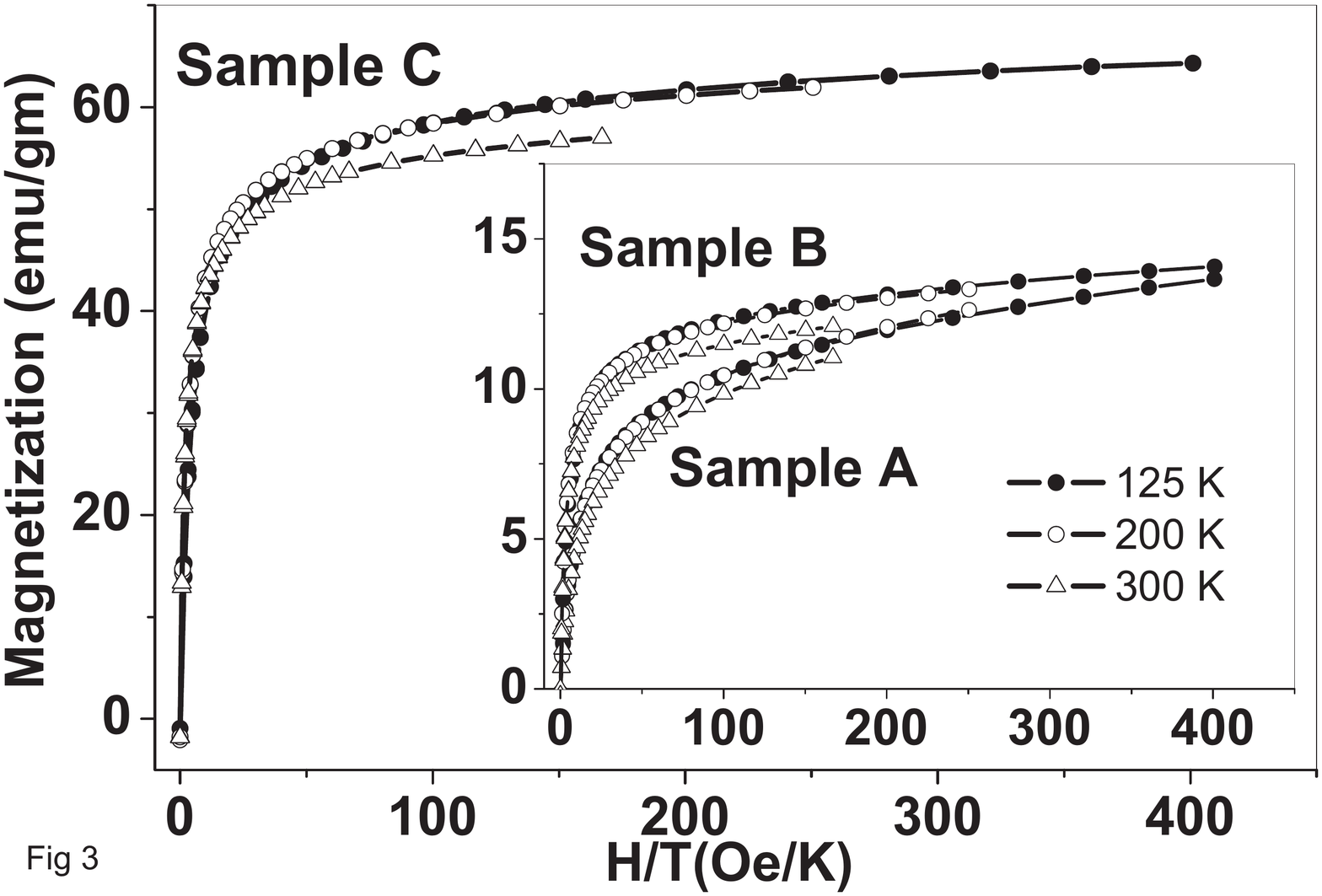}
\caption{Magnetization vs H/T for Sample A, Sample B and Sample C.}
\end{figure}

\begin{figure}
\includegraphics*[width=9.5cm]{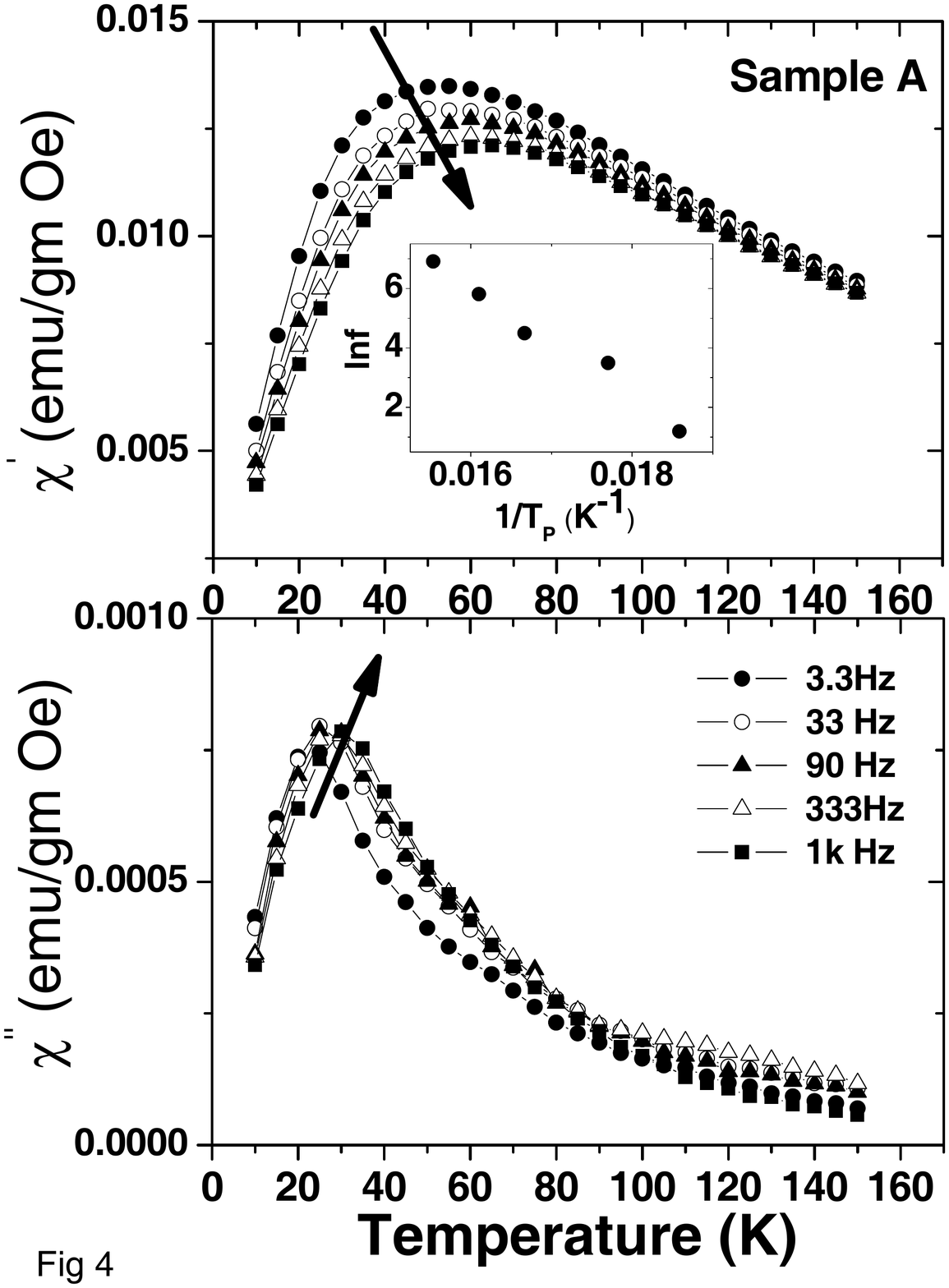}
\caption{$\chi^\prime$ and $\chi^{\prime\prime}$ for Sample A taken at different freqencies. The arrow indicates the shift in the peak temperature as a function of frequency. Inset shows the linear dependence of ln f vs 1/T$_{p}$. T$_p$ is the temperature where $\chi^\prime$ peaks.}
\end{figure}

\begin{figure}
\includegraphics*[width=9.5cm]{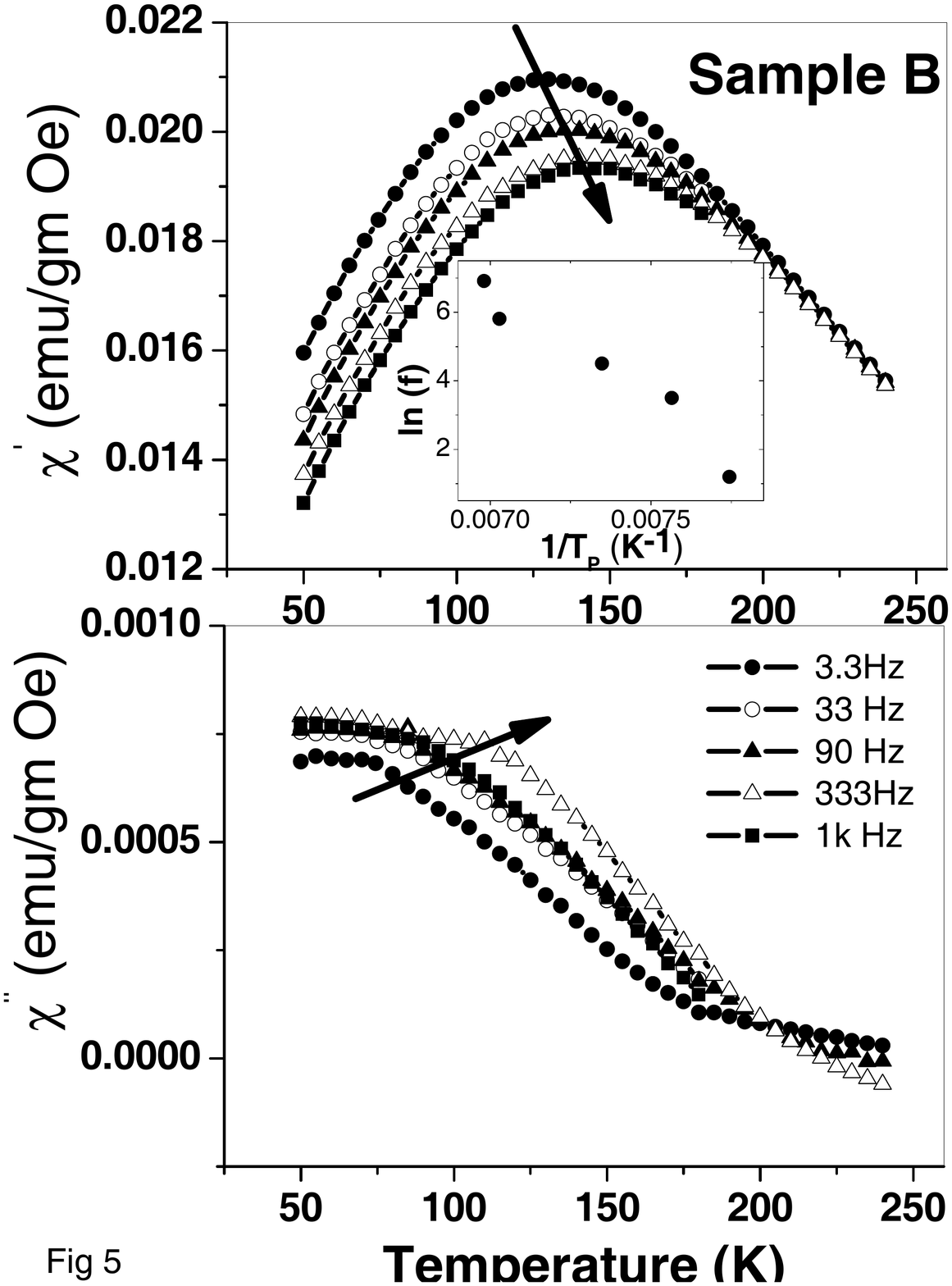}
\caption{$\chi^\prime$ and $\chi^{\prime\prime}$ for Sample B taken at different freqencies. The arrow indicates the shift in the peak temperature as a function of frequency. Inset shows the dependence of ln f vs 1/T$_{p}$. T$_p$ is the temperature where $\chi^\prime$ peaks.}
\end{figure}

\begin{figure}
\includegraphics*[width=9.5cm]{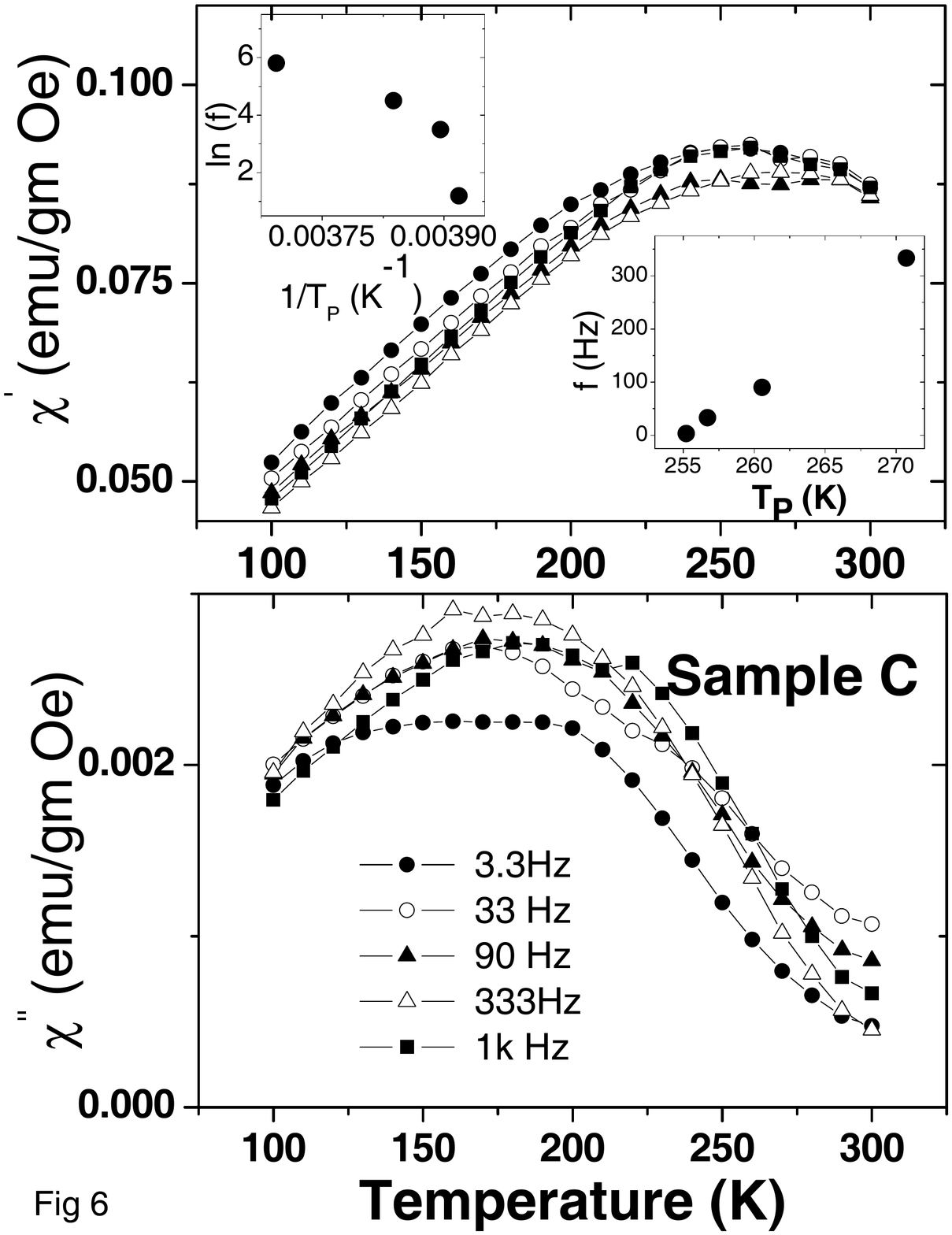}
\caption{$\chi^\prime$ and $\chi^{\prime\prime}$ for Sample C taken at different freqencies. There is no significant shift in the peak temperature as a function of frequency.Insets shows the ln f vs 1/T$_{p}$ and f vs T$_{p}$ obtained from the peak. T$_p$ is the temperature where $\chi^\prime$ peaks.}
\end{figure}

In figs~4, 5 and 6 we show the frequency dependence of the in-phase ($\chi^{\prime}$) and out-of-phase ($\chi^{\prime\prime}$) components of the ac susceptibility for the three samples. There are mainly three important features to be noted here. (1) Sample A and Sample B show a decrease in magnitude of peak values in $\chi^{\prime}$ with increase in applied frequency along with the peak shifting to higher temperatures. This feature is common in frustrated spin systems\cite{kleemann,suzuki} and shows the importance of interaction between the magnetic nanoparticles. For an assembly of independent single domain nanoparticles (canonical superparamagnets) this behaviour is not expected. The magnitude of $\chi^{\prime\prime}$ on the other hand shows small increase with frequency, again quite abnormal for a canonical superparamagnet. It is to be noted that Sample C does not show any significant frequency shift in the peak position in both the ($\chi^{\prime}$) and $\chi^{\prime\prime}$, (2) The temperature of the peak position of the $\chi^{\prime}$ and $\chi^{\prime\prime}$ increases with Au content and (3) The peak position of $\chi^{\prime}$ and $\chi^{\prime\prime}$  do not coincide and are far separated in temperatures. The separation between the peak positions of $\chi^{\prime}$ and $\chi^{\prime\prime}$, $\Delta T$ increases considerably with increase in Au content. 

Both superparamagnetic blocking behaviour (FC and ZFC separation) and super-spin glass behaviour can be studied by looking at the frequency ($f$) dependence of the temperature at which the real part of the AC susceptibility peaks (T$_p)$). The T$_p$ extracted from our data for the different frequencies are plotted in the insets of the real part of the susceptibility plot of the respective samples. For a simple superparamagnet, N\'{e}el-Brown-Arrhenius law holds, $f$=$f_0$exp[-E$_a$/$k_B$T$_p$],
where $f_0$ is the attempt frequency for coherent rotation of all spins within
a nanoparticle (superspin flip), E$_a$ is an activation energy.
Vogel-Fulcher law is a simple modification of this with, T$_p$ replaced by
T$_p$-T$_0$ where $k_B$T$_0$ is of the order of average near neighbour interaction
energy. Thus  $ln f$ will show a linear dependence with a negative
slope as a function of 1/T$_p$ which is observed in our pure (sample
A) and to some extent in low Au (sample B), indicating that they can be 
thought of as assembly of weakly interacting superparamagnetic
nanoparticles \cite{denardin}.

In the case of a spin glass state, the critical scaling law, indicates the existence of a frozen disordered state (collective), or something like a phase transition (in this case
a spin glass state)\cite{frus,djurberg,wandersman}. It is given by 1/$f$=1/$f_0$ [[T$_p$-T$_p$(0)]/T$_p$(0)]$^{-z\nu}$, where T$_p$(0) is the FC-ZFC bifurcation temperature, or the peak temperature for the lowest frequency used (DC limit) and $z\nu$ is a dynamical critical exponent.
If $z\nu$ is 1 then, one can obtain, $f$ proportional to T$_p$ which is observed in the case of the high Au sample (Sample C). For bulk ferromagnet/ferrimagnets\cite{chaikin}, $1.2<z\nu <2$ and for bulk 3-d spin glasses\cite{mydo}, $5<z\nu<11$. Theoretical result\cite{blundell} for
3-d Ising spin glass is, $z=5.85\pm 0.3$ and $\nu=0.29\pm 0.07$, giving a value of $z\nu \approx 1$ even though it was shown\cite{bern}, that this dynamical exponent is non-universal and even depend on temperatures. The power law dependence of frequency on peak temperature for sample C, as opposed to activated(exponential)dependence for samples A and B, shows that sample C has  spin glass like features.
 
\begin{figure}
\includegraphics*[width=9.5cm]{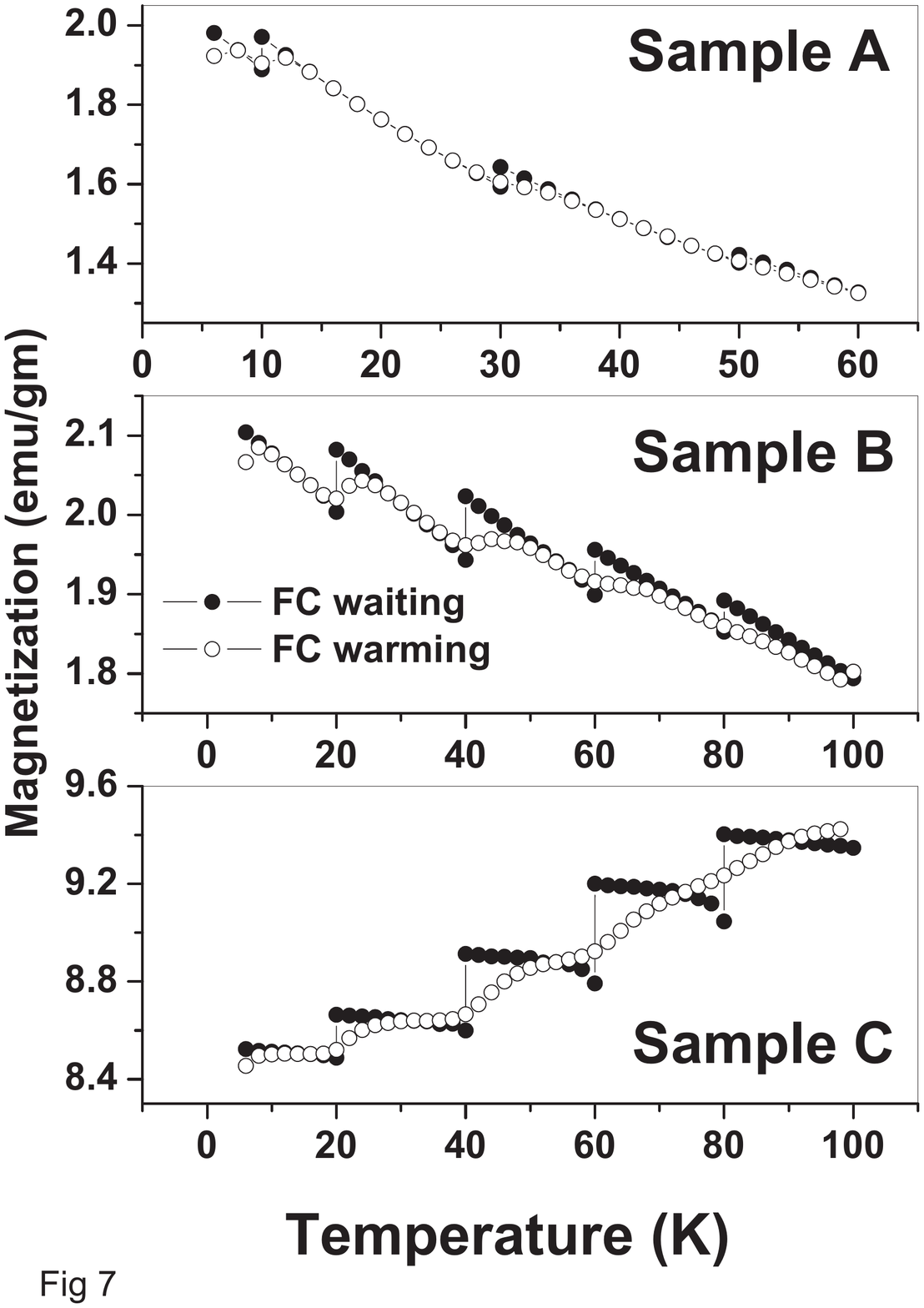}
\caption{Memory effect in the dc magnetization. The temperature at which there are steps in the FCC (field cooled cooling - black dots) magnetization data indicate the temperature at which the field was dropped to zero and measurement was stopped for 2 hours. The white dots correspond to the magnetization measured during continuous warming in the presence of the same field.}
\end{figure}

In fig~7 we show the magnetic memory effect in FC magnetization for all the three samples using the protocol discussed earlier. The temperature at which there are steps in the FCC (field cooled cooling - black dots) magnetization data indicate the temperature at which the field was dropped to zero and measurement was stopped for 2 hours. The white dots correspond to the magnetization measured during continuous warming in the presence of the same field. We see a clear signature of the memory effect at the same halt temperatures. It is clear from the plots that the memory effect 
is more pronounced in the sample with Au (Sample B and Sample C) and
nearly absent in pure Fe$_3$O$_4$ sample. Memory effect in the
magnetization is known to occur both in spin glass systems as well as
in superparamagnets with a varied size distributions or
polydispersity\cite{duttagupta1,duttagupta2}. As can be seen from the
TEM micrographs, our magnetic particles Fe$_3$O$_4$ do not have much
size variations. Hence, sample A which is pure magnetic Fe$_3$O$_4$,
do not show much memory effect indicating no polydispersity in the magnetic particle size. In other two samples, the magnetic particle size distribution does not change, but the particle size of the Au changes. Since the memory effect is stronger in sample C than in Sample B, which indicates that with increase in Au content the system slowly evolves from a superparamagnet to a superspin glass system. To get a better understanding we measured thermoremanent magnetization and the zero field cooled magnetization relaxation (ZFCM) versus time for the nanocomposite samples (Sample B and Sample C).

\begin{figure}
\includegraphics*[width=9.5cm]{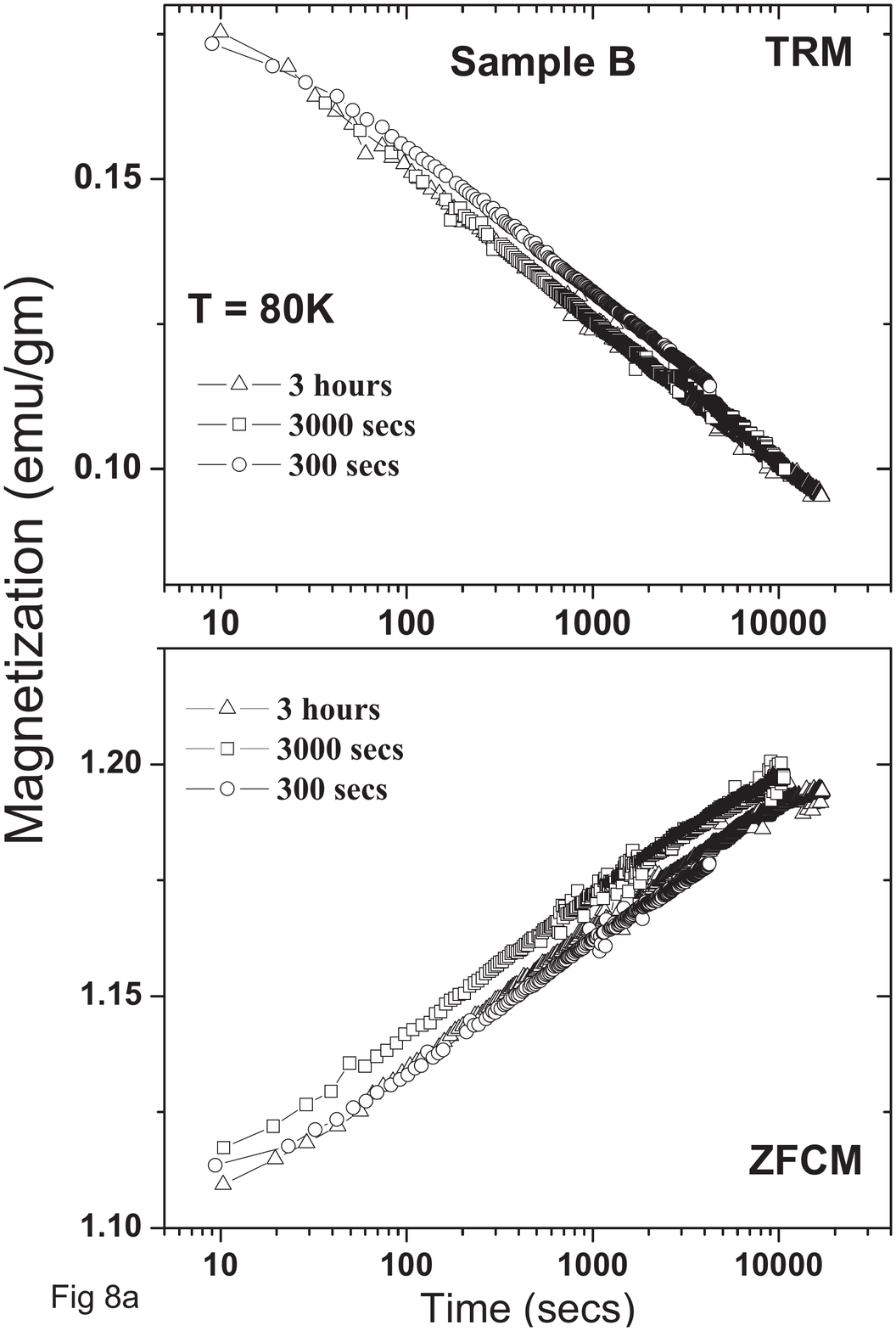}
\end{figure}

\begin{figure}
\includegraphics*[width=9.5cm]{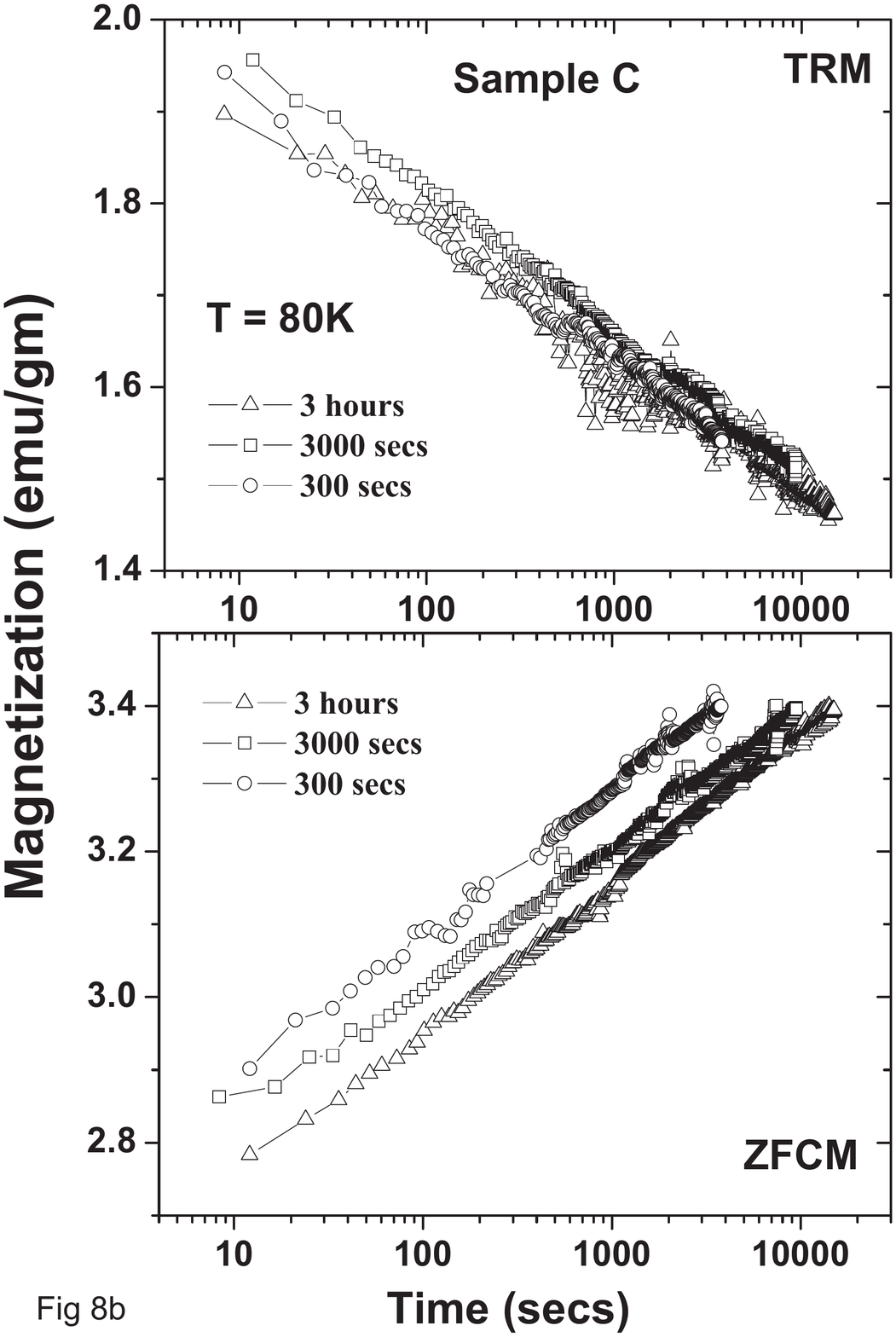}
\caption{Thermoremanent magnetization relaxation (TRM) and the zero field cooled magnetization relaxation (ZFCM) for (a) Sample B  and (b) Sample C taken at 80K with different wait times of 300 secs, 3000 secs and 3 hours. Sample B shows a weak wait time dependence on the relaxation whereas Sample C shows a significant wait time dependence on the relaxation.}
\end{figure}

In fig~8(a,b) we have plotted the thermoremanent magnetization relaxation (TRM) and the zero field cooled magnetization relaxation (ZFCM) for Sample B  and Sample C taken at 80K. Two basic features are observed. (1) The magnetic relaxation is slow, logarithmic in time (indicating that it is a glassy state), which can be clearly seen as a linear plot in a semi-log scale for both the TRM and ZFCM, (2) For both the samples, the TRM does not show a distinct wait time dependence, whereas ZFCM shows a wait time dependence which becomes more and more pronounced with the increase in Au content, longer the $t_w$, the slower is the relaxation (aging effect). Wait time dependence of TRM and ZFCM is a typical property of spin glasses and not of superparamagnets \cite {Nam,wandersman}. Both increase of memory effect in FC magnetization and increase in wait time dependence of magnetic relaxation increases with increase in Au content points at a slow evolution towards a spin glass state from a weakly interacting superparamagnetic state, as Au content is increased.

\begin{figure}
\includegraphics*[width=9.5cm]{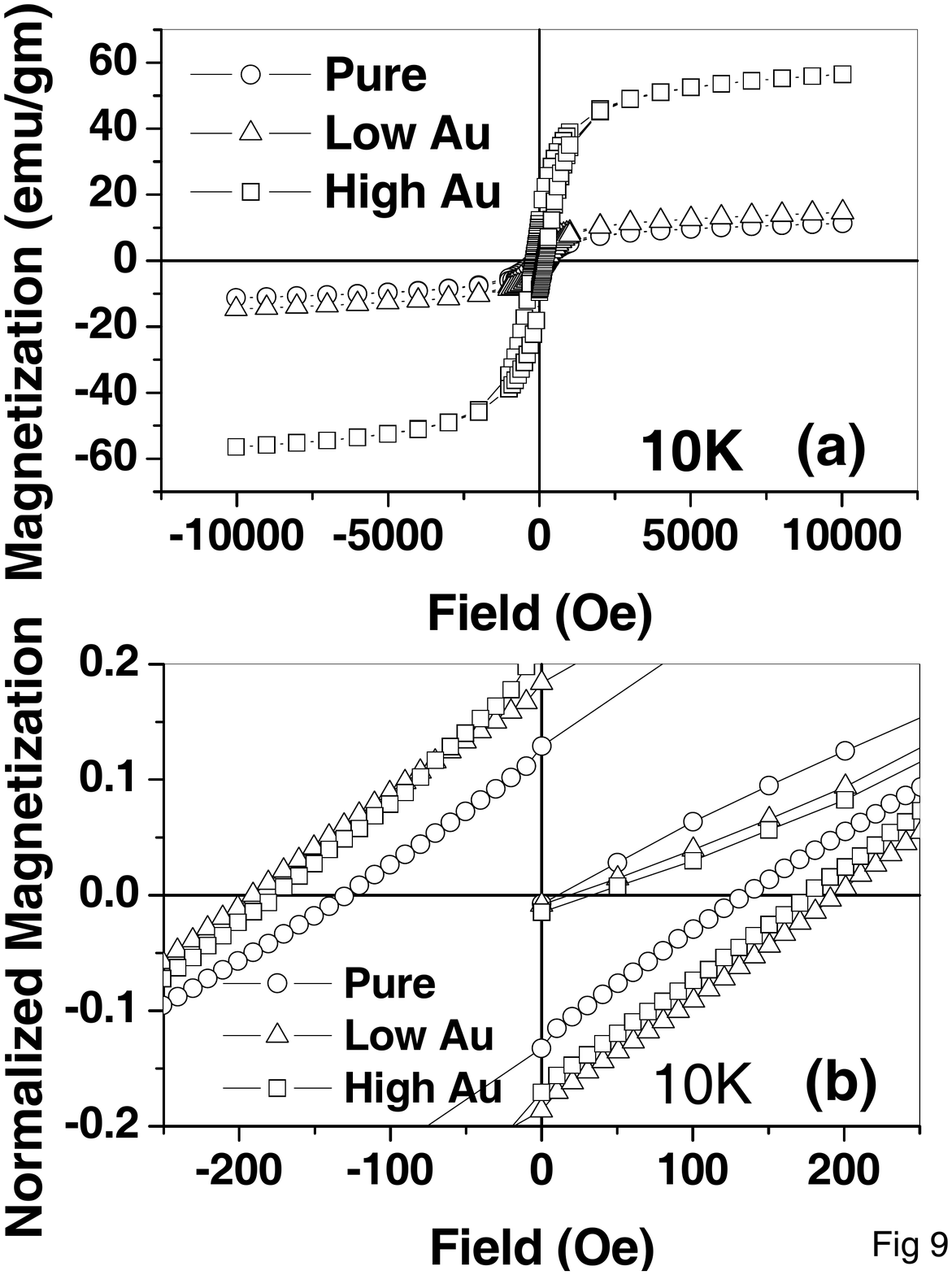}
\caption{(a) ZFC hysteresis loop of the samples taken at 10 K. (b) Plot of the
normalised magnetization (M/M$_{sat}$ in the expanded scale, where M$_{sat}$ is the saturation magnetization)}
\end{figure}

\begin{figure}
\includegraphics*[width=9.5cm]{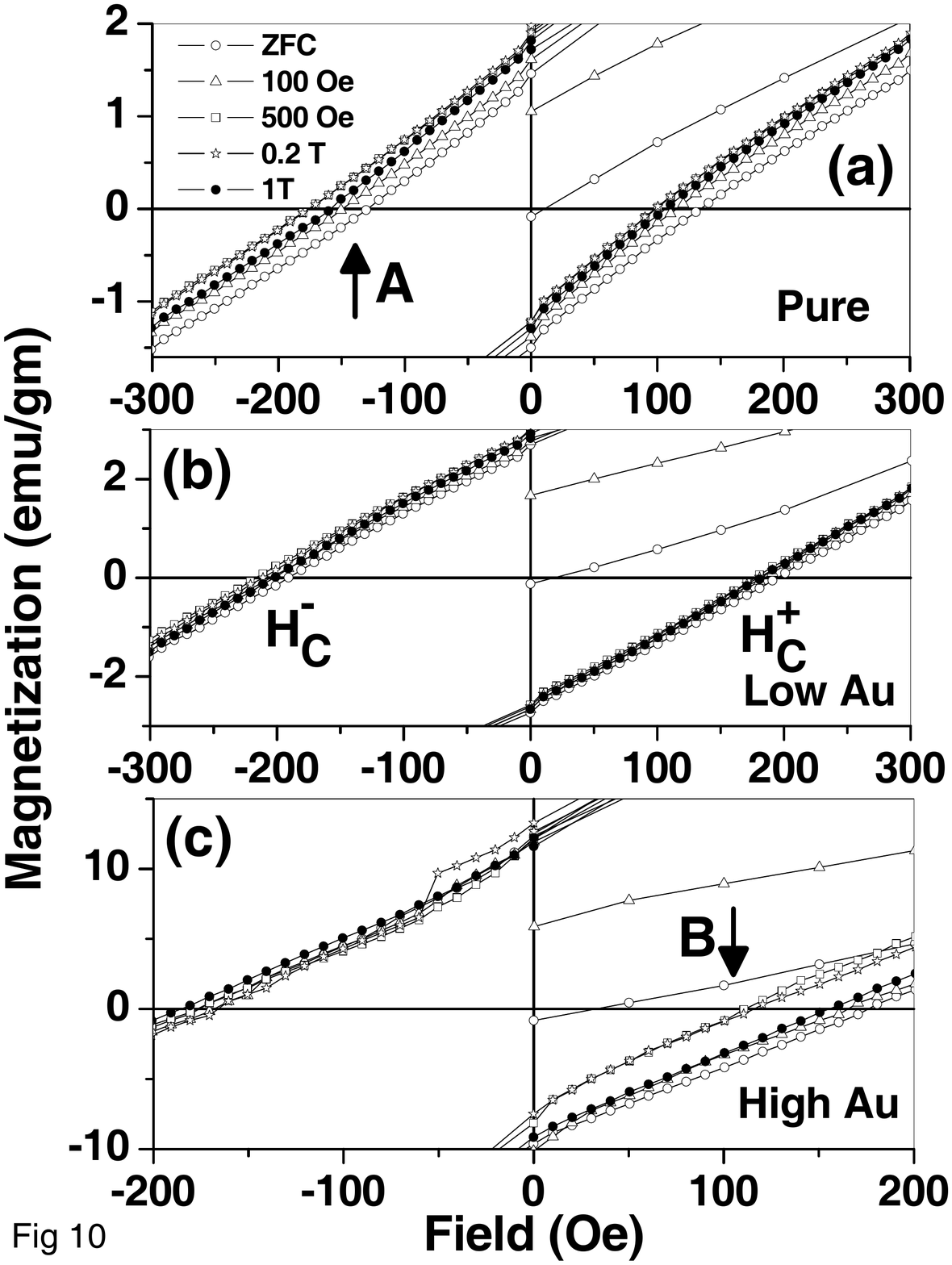}
\caption{The magnetic hysteresis results taken as a function of cooling field at 100 Oe, 500 Oe, 0.2 T and 1 T, along with the ZFC hysteresis for all the three samples. The arrow A shows the spread out of H$_c^-$ for pure sample. Arrow B indicates the spread out of H$_c^+$ for high Au sample.}
\end{figure}

To further see the slow evolution of magnetic property in our system, we have measured the exchange bias for all the three samples. Exchange bias is a powerful measure to study phase seperated magnetic anisotropies in a system. Fig~9(a) shows the ZFC hysteresis loop of
the samples taken at 10 K. We can clearly see that the high Au sample shows the largest saturation magnetization. To compare the hysteresis we plot the normalised magnetization (M/M$_{sat}$, where M$_{sat}$ is the saturation magnetization) in fig~9(b). The coercivity initially increases with low Au concentration but again decreases for the high Au content sample. This interesting observation prompted us to measure the FC magnetization hysteresis on the samples to detect existence of exchange bias. Fig~10 shows the magnetic hysteresis results taken as a function of cooling field at 100 Oe, 500 Oe, 0.2 T and 1 T, along with the ZFC hysteresis. When the coercive field in the negative field direction H$_c^-$ is not equal to the coercive field in the positive field direction H$_c^+$, the system is said to exhibit exchange bias. Generally it is observed that the H$_c^-$ are more spread out than H$_c^+$ for typical nanoparticle systems \cite{bianco,he,kavich,nogues}. Also interestingly even this kind of behavior is observed (i.e., the nature of the spread out) as a function of pressure \cite{wang} and as a function of nanoparticle size \cite{eftaxias}. In fig~10 we observe that as the Au content increases, the system shows a change over of the spread out of H$_c^-$ for pure sample (marked by arrow A) to spread out of H$_c^+$ for high Au sample (marked by arrow B). In fig~11 we show the exchange bias (H$_E$ = (H$_c^+$+H$_c^-$)/2)), coercivity field  (H$_C$ = (H$_c^+$-H$_c^-$)/2)), 
remanence assymetry (M$_E$ = (M$_+$+M$_-$)/2)) and magnetic coercivity (M$_C$ = (M$_+$-M$_+$)/2)) where M$_+$ and M$_-$ are the positive and the negative remanant magnetizations.
 It has been shown that M$_E$/M$_{sat}$ $\propto$ -H$_E$ \cite{solamon} for systems exhibiting exchange bias. We observe a non-monotonic behaviour of 
exchange bias as a function of cooling field as shown in fig~11(a). The exchange bias increases sharply at low cooling fields and then decreases for higher cooling fields. A similar behaviour is also observed in the normalised remanant assymetry (M$_E$/M$_{sat}$) as shown in fig~11(c) following the relation above. This sharp increase of the exchange bias at low cooling field and then a decrease at higher cooling fields is also observed on other nanoparticle systems \cite{bianco,kavich}. Another important feature to notice is that the exchange bias and the normalised remanant assymetry initially decreases on addition of Au and then increases for the high Au content sample. On the other hand, the pure sample has the lowest coercive field. Coercivity increases sharply with increase in Au content, but decreases again for high Au sample. The cooling field dependence of coercivity of sample A and B
are very similar, while sample C shows markedly different variation
with cooling field.
We shall discuss the detailed dependence  of
exchange bias and coercivity on cooling field later.

\begin{figure}
\includegraphics*[width=9.5cm]{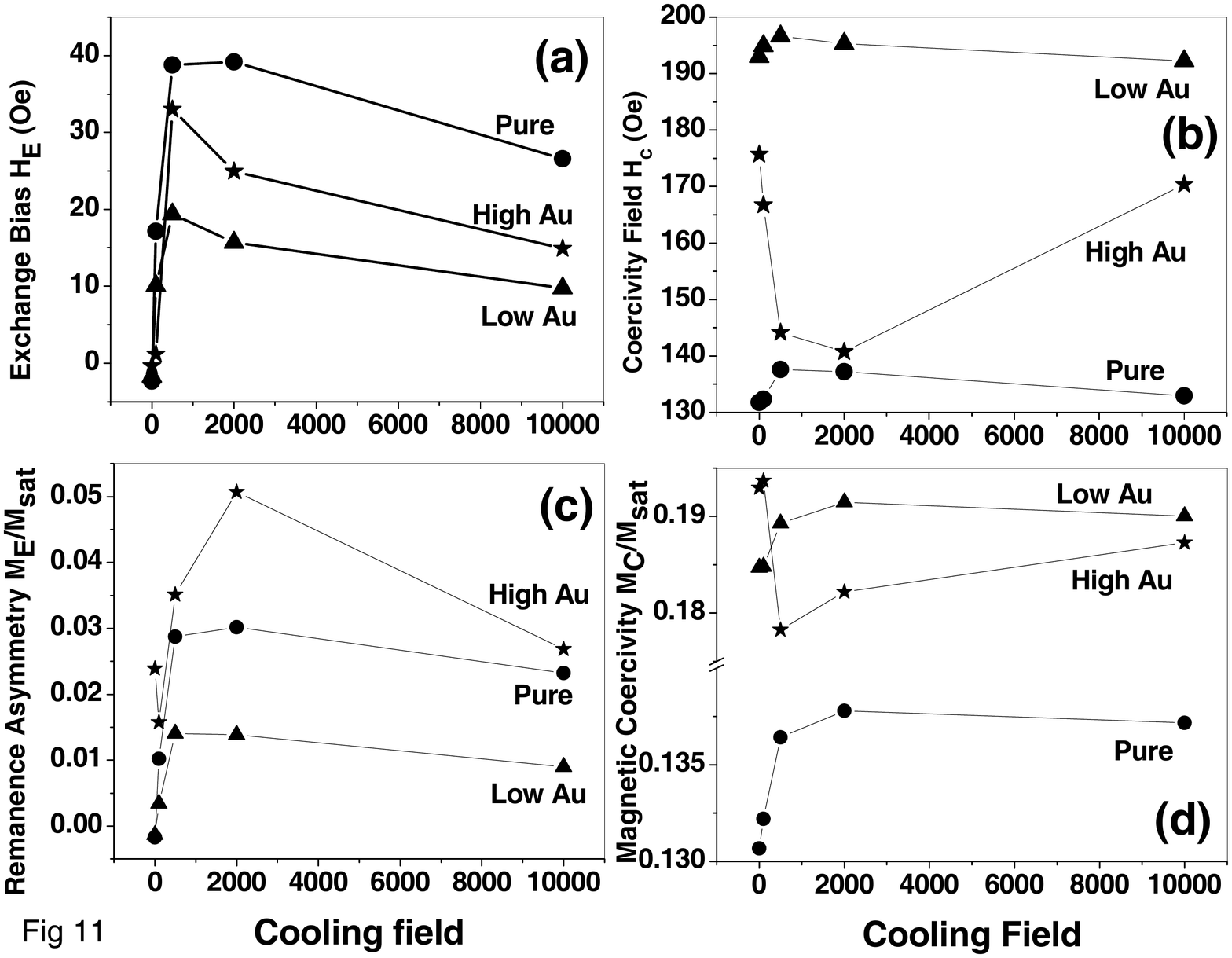}
\caption{(a) Exchange bias (H$_E$ = (H$_c^+$+H$_c^-$)/2)), (b) Coercivity field  (H$_C$ = (H$_c^+$-H$_c^-$)/2)), (c) Remanence assymetry (M$_E$ = (M$_+$+M$_-$)/2)) and (d) Magnetic coercivity (M$_C$ = (M$_+$-M$_+$)/2)) where M$_+$ and M$_-$ are the positive and the negative remanant magnetizations as a function of the cooling field for all the three samples.}
\end{figure}

First we discuss the curious increase in magnetization and increase in blocking temperatures
with increase in Au particle size. As we have emphasized before that the Fe$_3$O$_4$ nanoparticles 
in our sample have a very narrow size distribution as seen from the TEM images and the absence of memory effect. The difference between the three samples is the variation in Au content. The interesting thing is that with increase in Au content the magnetic moment/gm is increasing. This behaviour we believe is new and unexpected. Typically Fe$_3$O$_4$ nanoparticle moment density is much lower than that of bulk Fe$_3$O$_4$ \cite{chen2}. This is generally thought to be due to finite size effect and surface spin canting due to lower coordination number and strain or structural deformation at the surface\cite{surf}. It has been observed that in Au coated  Fe$_3$O$_4$ nanoparticles the moment reduces further\cite{srikanth1,wang}, indicating that surface moments might be further disordered due to interaction with Au electrons, leading to the reduced moment. Our system Au-Fe$_3$O$_4$  is very different from such a core-shell structure, though we have large number of interfaces between Au and Fe$_3$O$_4$ on the surface. In any case the disordering of canted moments on Fe$_3$O$_4$ surfaces due to conduction electrons of Au should still be occurring and hence an increase of net moment is rather surprising. We also see a progressive and systematic increase of blocking temperature with increase in Au content. A simple guess would be that Fe$_3$O$_4$ is spin polarising Au very close to the interface. This will increase the net moment as well as increase the effective volume of the magnetic nanoparticles. Since blocking temperature for larger volume particles is higher, this might be consistent with larger blocking temperature as well as larger moment for samples with higher Au content. Spin polarization of nonmagnetic metals in contact with ferromagnets
 was studied extensively by Hauser \cite{hauser} experimentally and theoretically by Clogston \cite{clog}. They found that the spin polarization can at best penetrate a length scale of  
1-2 nm in a nonmagnetic metal in contact with a ferromagnet. This is rather small to explain the large change in blocking temperature coming from an effective volume increase of magnetic nanoparticle due to spin polarization of Au electrons near the interface. Since spin polarization does not extend to large distances, it cannot explain the continued increase in net moment and blocking temperatures with increase in Au content. Experimentally magnetic moment of Au near Co/Au interface has been measured from magnetic X-ray circular dichroism \cite{coau} to be about 0.062 $\mu_B$ per Au atom near the interface. The origin is the spin-orbit splitting of Au surface states (inversion symmetry is lost on the surface), but the moments are far too small to explain our observed increase in moment in Au- Fe$_3$O$_4$ system. So we have to look elsewhere to explain this phenomena.

A set of interesting experimental results on the magnetic properties of some nanostructures has been recently published. Large magnetic moments were detected on the surface layers of thin films of borides and oxides\cite{oxides,borides}. Ferromagnetic hysteresis at room temperature was
 measured in Au nanoparticles\cite{Au} and Au nanoparticles/films with thiol patches 
on top\cite{authiol,carmeli}. Spin splitting of surface electronic states was observed in Au(111)\cite{lashell}, and Bi\cite{koroteev}. Similar magnetism was detected in Pd nanoparticles also\cite{sampedro}. A common characteristics of all of these unusual magnetic behavior seems to be that local anisotropy is very large compared to typical anisotropy strengths of  well known harder materials. Recently a theoretical attempt was made by Hernando et. al.\cite{hernando} to explain magnetic moment in Au with thiol patches on top. Important difference of our system is that here we have an interface between Au and (magnetic) Fe$_3$O$_4$ unlike Au and (nonmetallic) thiol. As we shall see, this has significant consequences. We shall assume the existence of a contact potential $U$ and a radial electric field (perpendicular to the interface) $E=-(dU/dr)_{r=\eta}$ at the Au - Fe$_3$O$_4$ interface. Free electrons of Au can be captured in large atomic like bound orbitals of circumference $\eta$ at the domain boundary potential step. With the spin component of the bound Au electron along the $z$ axis being $s_z$, the Hamiltonian for these bound electrons can be written down as

\begin{equation}
H= \frac{\hbar^2 L_z^2}{2 m \eta^2} -\alpha \hbar^2 L_zs_z  + \lambda s \cdot \sum_i^{\eta/a} S_i
\end{equation}

\noindent Here $\eta$ is the length of the interface of any Fe$_3$O$_4$ particle with Au, $\alpha$ is the spin-orbit coupling strength and is proportional to the gradient of the contact potential, and $\lambda$ is the exchange (antiferromagnetic\cite{clog}) coupling strength of the Au electron having spin $s$ and any Fe moment at the boundary, $i$ being site index of the Fe moments along the interface having a spin $S_i$ and $a$ is the average Fe-Fe distance in the Fe$_3$O$_4$ particles. We choose $\eta= 5$ nm, $\alpha \hbar^2 = 0.4$ eV, a value estimated from experimental spin splitting observed on Au surfaces\cite{lashell} and $\lambda =0.1 $ eV (typical values of contact exchange interaction) \cite{clog} for illustrative purposes and write the Hamiltonian as

\begin{equation}
 H= {\hbar^2 L_z^2 \over 2 m \eta^2} -\alpha \hbar^2 L_zs_z  + \lambda \eta s_z M_z +\lambda \sum_{\eta} {1\over 2}(S_i^+s^- +S_i^-s^+)
\end{equation}

\noindent where $M_z$ is the average z component moment of the surface Fe atoms. The last part is the transverse part of the contact exchange interaction, that gives rise to spin flip scattering between the boundary Fe moments and the Au electrons (both bound and free electrons). Forgetting the last term for the time being, we find that when $M_z=0$ the energy is negative for $L_z= 1$ to 160, i.e., one could have 160 electrons filling such bound orbitals all with same $s_z$. In fig~12 we have plotted the energy versus $L_z$ values for different values of $M_z$ (average boundary Fe moment). We can see, that for $M_z >1.0$ there are no bound states (negative energy) at all, for the chosen values of parameters. In other words if the $z$ component of the boundary spins add upto large values then it is not possible to have bound Au electrons along the interface with large orbital angular momenta. On the other hand when average $M_z=0$ like in Au-thiol (nonmagnetic) interface, it is possible to have large number of bound  states occupied with electrons (having $L_z$ values from 1 to large values, and same $s_z$ to minimise exchange part of the coulomb correlation energy) near the interface, giving a large net moment. The spin flip scattering (the last term in eqn.2) by the free as well as bound Au electrons with the boundary Fe moments, on the other hand try to randomize the Fe moments giving rise to lesser $M_z$ value. Thus, samples with larger Au concentration will have larger concentration of free electrons, and hence reduces the average boundary Fe moments more efficiently compared to sample with lesser Au/free electron concentration. This could be the reason why, we find larger moment in samples with larger concentration of Au. If we had for example, a composite of Fe$_3$O$_4$ and any nonmagnetic insulating particles, then our mechanism does not allow the existence of large orbital magnetic moments. Since the additional magnetic moments come from orbital moments, this implies a high magnetic anisotropy of the Fe$_3$O$_4$ + Au bound electrons composites. This increased magnetic anisotropy of the elementary super moments as well as slightly increased effective size (due to additional bound electron states near interface) could be responsible for the huge increase in the blocking temperature of the Fe$_3$O$_4$ - high Au nanocomposite sample. The frequency dependence of both real and imaginary part of the magnetic susceptibilty of the high Au content sample shows very little frequency dependence compared to  pure Fe$_3$O$_4$ or sample with small Au content. This is because the effective supermoments are very large with an enhanced anisotropy energy barrier. Another important observation is the difference in FC magnetization below the blocking temperatures for all samples. The individual (non-interacting) blocking model predicts a monotonous increase in FC magnetization for superparamagnets, i.e the FC magnetization increases with decreasing temperatures till all the particles are blocked. This is what is happening for the pure Fe$_3$O$_4$ particles. The dipole-dipole interaction which should be greatest in this sample is not strong enough to induce much spin glassiness at low temperatures. The sample with large Au content on the other hand shows large memory effect, noticeable wait time dependence of TRM and ZFCM, as well as FC magnetization saturation below the blocking temperatures. This shows that the spin glass state in high Au content sample arises from the RKKY type interaction between the surface spins of oxide nanoparticles mediated by the Au electrons and not due to the dipole-dipole interaction between the magnetic particles. In Co-Ag granular composite films\cite{du}, it is found that memory effect weakens with increase in the volume fraction of magnetic (Co) clusters, a result very similar to ours.

\begin{figure}
\includegraphics*[width=9.5cm]{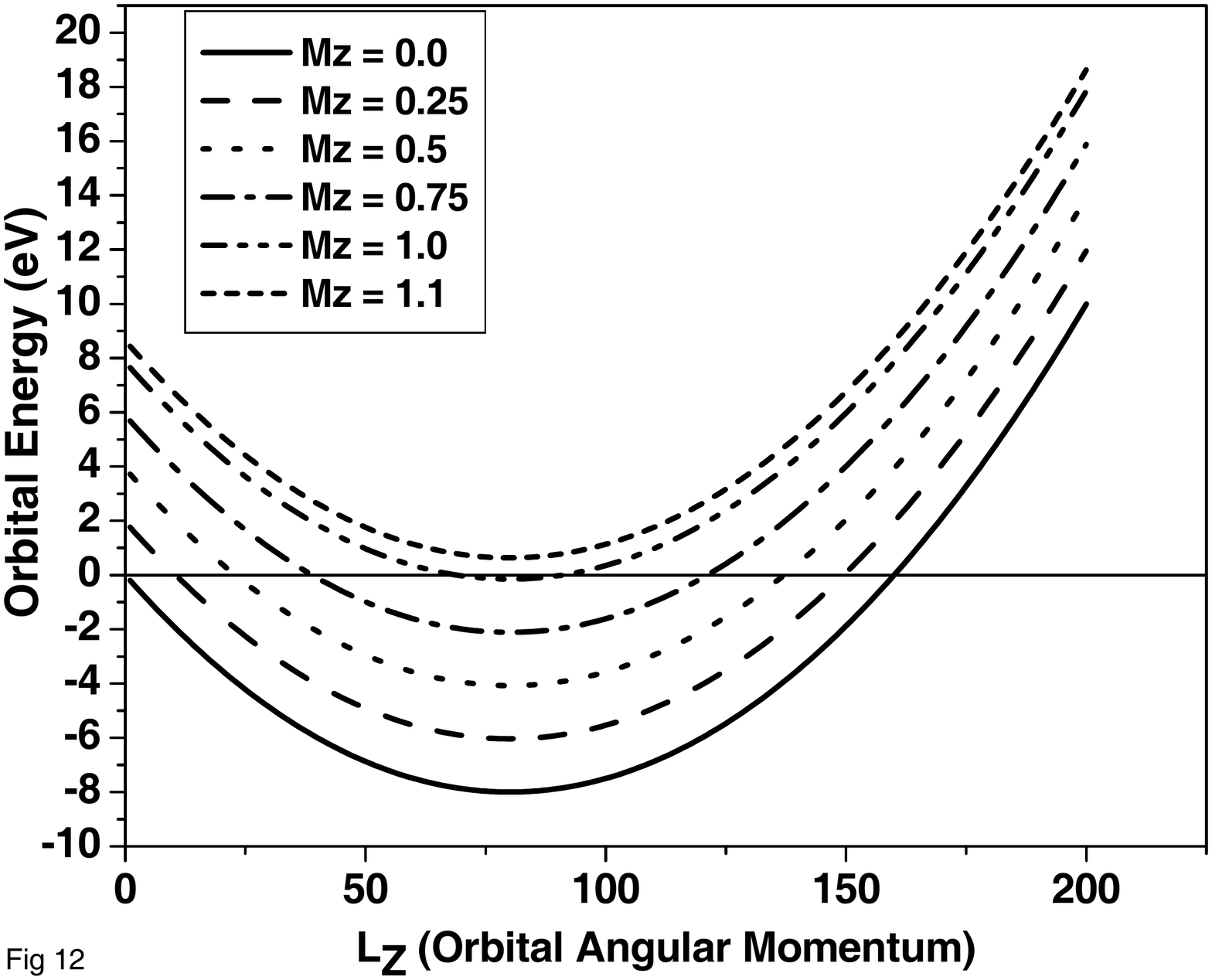}
\caption{Orbital Energy versus $L_z$ values for different values of $M_z$. We can see, that for $M_Z >1.0$ there are no bound states (negative energy).}
\end{figure}

Though many groups have worked with Fe$_3$O$_4$-Au nanoparticle composites, to our knowledge there has been no report so far on such large enhancement of net magnetization of the composite. There could be several reasons for that. (1) Since that $z$ axis should be very well defined throughout the Au-magnetic particle interface. The value of effective $\eta$ is  very small
for very small sized Au nanoparticles, or for interfaces where the plane enclosed by $\eta$ deviates from a plane too much. For example in thiol capped Au nanoparticle system the moment/Au atom is very large in thin films compared to small nanoparticles\cite{authiol}. (2) on the other hand if the Au particles are large, then the core diamagnetism of Au electrons may cancel out the large orbital moments at the interface. Thus, there seems to be an optimum size of the Au particle which will show maximum magnetization, beyond which the diamagnetic term will dominate. In our case the base material (Fe$_3$O$_4$) itself is magnetic and Au-diamagnetism is very small compared to the magnetic base material.

The existence of exchange bias, shows that there is a second component
of magnetisation in Fe$_3$O$_4$ nanoparticles that do not coherently
flip or rotate like the core spins, when the external magnetic field
is cycled, thereby  giving the core spins an average uniaxial anisotropy.
 Exchange bias is routinely observed in magnetic
nanoparticles, and is usually ascribed to the surface spin glassiness
of the magnetic particles. For magnetic nanoparticles with surface spin
glasses, as is usual for nanoparticles of size 20 nm or less, the
exchange bias field
usually rises with cooling field, goes through a peak and comes down
at larger cooling field \cite{bianco,kavich}. This is because, the clamped magnetic field
(pinned to the lattice and has slow evolution with time) has a similar
variation with cooling field. The clamped magnetic field at any given temperature is the difference between the ZFC magnetization and the FC magnetization which dependes on the applied cooling field. This clamped magnetic field is non-zero and has very slow time evolution within the time of measurement, and is characteristic of spin glasses which also give rise to wait time dependence in the relaxation measurement (ZFCM and TRM). The reduction of
clamped magnetisation at higher cooling field is to be expected on the
physical ground that spin glass state will be ultimately destroyed
under high magnetic field. The variation of exchange bias field of pure Fe$_3$O$_4$
particles versus the cooling field has this peaking effect at
intermediate  cooling field.
Both low Au and High Au samples has also the same features. But the
absolute values of exchange bias fields of low Au sample are smaller
than that of the high Au and pure Fe$_3$O$_4$ samples. This needs some discussion.

In pure Fe$_3$O$_4$ the interaction between the particles are dipolar
in nature. This interaction is between the large superparamagnetic
moments (core + shell combined moments). With addition of Au 
there are additional RKKY type of interaction between only the shell moments of Fe$_3$O$_4$ particles of the type $ H= J_0 {\cos (2k_F r+ \phi) \over r^3} e^{-{r\over \lambda}} $ where $J_0$ is the interaction strength and $\lambda$ is the mean free path of Au conduction electrons. One has to also distinguish between interface area $\eta < \eta_c$ and $\eta > \eta_c$. Large orbital moments can arise only around larger particles (because of lesser kinetic
energy - see Eq.1). With incorporation of Au, the surface moments of the smaller particles  will be driven to further glassiness by the
RKKY interaction within a characteristic length scale of $r<\lambda$. Because of this enhanced glassiness (frustration) the cooling field
wou;d not be able to induce as large a surface magnetisation as it was possible for pure Fe$_3$O$_4$ particle surface spins. This is probably why the
exchange bias field values are much smaller in low Au+Fe$_3$O$_4$ particles. Another consequence of this spread of spin glassiness over a larger
length scale is the large increase in coersive field for the low Au samples, because it is difficult to reverse total magnetisation when
glassy correlation between surface moments is of longer range. The large orbital moments on the  larger size particles on the other
hand cannot be flipped easily by the conduction electrons because of large local Hunds exchange energy cost. They increase the net
saturation moments.

For high Au sample several interesting things might be happening (1) larger percentage of Fe$_3$O$_4$ particle now have large orbital moments around them increasing the net saturation magnetisation as well as the blocking temperatures and (2) the mean free path $\lambda$ is now much more than that of the low Au samples and hence glassy spin correlations is of much longer range. But since the surviving smaller particles are now far separated, the glassy spin corelation is much weaker (lesser $J_0$) in strength. That is why the cooling field can easily induce larger surface magnetisation and hence larger exchange bias field on the core moments of the smaller particles. The core moments of the larger particles also feel some biasing from
the orbital moments because now they are truely large in magnitude.
This is why the exchange bias field versus cooling field curve for the
higher Au sample is consistently above that of the low Au sample for
all cooling fields. With increase in cooling field, both surface magnetisation and large
orbital moments increases and since the frustrated RKKY interaction
strength is small (even though long range), it is easier to rotate the
net magnetization coherently by small magnetic fields. This is the reason the why
coersive field falls initially with cooling field for smaller fields. The unusual increase of coercivity at larger cooling field comes as a surprise, and we can only speculate at this stage
of our work that at  high cooling fields, the large orbital
moments might induce ferro-spin polarisation near them and suppress the
spin flip rate (between surface moments and conduction electrons)
 considerably. RKKY interaction strength is quadratic (second order) in spin flip scattering rate and this gets reduced at higher cooling fields. This effectively reduces 
communications between the Fe$_3$O$_4$ particles. System fragments into 
many rigid domains. This is possibly why we see a slow increase in coersive fields
at larger cooling fields for high Au samples.

\section{Conclusion}
In the present investigation we have observed that the
ensemble of pure Fe$_3$O$_4$ nanoparticles sample behaves superparamagnet like, with some surface spin glassiness, and it shows
negligible magnetic memory effect since the particle size distribution is nearly monodispersed. We noticed via magnetic measurements that on incorporation of gold (Au) nanoparticles the nanocomposite system slowly evolves from superparamagnetic to superspin glass with the increase of the Au nanoparticle size/content. The frequency dependence of the real part of the magnetic susceptibilty of the high Au content samples shows power law dependence as opposed to the pure Fe$_3$O$_4$ or sample with small Au content both showing activated (exponential) dependence. We observe that with the increase of Au nanoparticle size/content the nanocomposite exhibits strong memory effect, aging phenomenon and systematic evolution of exchange bias. This strong memory, aging effect and anomolous behaviour of exchange bias has been attributed to superspin glass nature of the nanocomposite. The most important observation in this study is the observation of enhancement of magnetization value upon incorporation of Au nanoparticles with Fe$_3$O$_4$ nanoparticles. The enhancement increases with the increase in the Au nanoparticle size/content. This phenomenon could be explained by modifying the Hamiltonian proposed by Hernando et. al. \cite{hernando} by including the exchange antiferromagnetic coupling of the interface magnetic moments of the magnetic nanoparticle with that of the conduction electrons of the Au nanoparticle for our case. From the present investigation it appears that anomolous magnetic property observed in the case of Au-magnetic or Au-nonmagnetic nanocomposites arises due to surface/interfacial effect. \\

{\bf Acknowledgement} The authors thank the TEM facility at Institute of Physics, Bhubaneswar for the TEM measurements.


\begin{thebibliography}{99}
\bibitem{bedanta}S.Bedanta and W.Kleemann, J.Phys.D.Appl. Phys. {\bf42}, 013001 (2009)
\bibitem{dormann1}J. L. Dormann, D. Fiorani, E. Tronc, Adv. Chem. Phys. {\bf 98}, 283 (1997)
\bibitem{dormann2}J. L. Dormann, L. Bessais D. Fiorani,J.Phys.C:Solid State Phys. {\bf21}, 2015 (1988)
\bibitem{freitas}P.P.Freitas and H.A.Ferreira, Handbook of Magnetism and Magnetic Materails, {\bf 4}:Novel Materails, ed H.Kronmuller and S.P.S.Parkin (New York, Wiley), 2507 (2007)
\bibitem{neel}L. N\'{e}e, Ann. Geophys.(C.N.R.S) {\bf5}, 99 (1949); Adv. Phys. {\bf4},191 (1955)
\bibitem{brown} W. F. Brown Jr., Phys. Rev. {\bf130}, 1677 (1963)
\bibitem{cullity} B. D. Cullity and C. D. Graham, Introduction to Magnetic Materaials,John Wiley and Sons, Inc., Hoboken, New Jersey, {\bf Chap 11}, 359 (2009)
\bibitem{kleemann}W. Kleemann, O. Petracic, Ch. Binek, G. N. Kakazei, Yu. G. Pogorelov, J. B. Sousa, S. Cardoso, and P. P. Freitas, Phys. Rev. B {\bf63}, 134423 (2001)
\bibitem{sun}Y. Sun, M. B. Salamon, K. Garnier and R. S. Averback, Phys. Rev. Letts., {\bf91} 16702 (2003)
\bibitem{sasaki} M. Sasaki, P. E. Jönsson, H. Takayama, and H. Mamiya, Phys. Rev. B {\bf71}, 104405 (2005)
\bibitem{suzuki} M. Suzuki, S . I. Fullem, I. S. Suzuki, L. Wang, and C-J. Zhong
Phys. Rev. B {\bf79}, 024418 (2009) 
\bibitem{tsoi} G. M. Tsoi, L. E. Wenger, U. Senaratne, R. J. Tackett, E. C. Buc, R. Naik, P. P. Vaishnava, and V. Naik, Phys. Rev. B {\bf 72}, 014445 (2005)
\bibitem{sahoo}S. Sahoo, O. Petracic, W. Kleemann, P. Nordblad, S. Cardoso, and P. P. Freitas, Phys. Rev. B {\bf67}, 214422 (2003)
\bibitem{chen} Xi Chen, S. Bedanta, O. Petracic, W. Kleemann, S. Sahoo, S. Cardoso, and P. P. Freitas Phys. Rev. B {\bf72}, 214436 (2005)
\bibitem{frus} D. N. H. Nam, R. Mathieu, P. Nordblad, N. V. Khiem and N. X. Phuc, Phys. Rev. B {\bf62}, 1027(2000), K. De, M. Patra, S. Majumdar and S. Giri, J. Phys. D: App. Phys. {\bf40}, 7614 (2007)
\bibitem{giri1} M. Thakur, M. Patra, S. Majumdar and S. Giri, J. appl. Phys. {\bf105} 073905 (2009)
\bibitem{duttagupta1}S. Chakraverty, M. Bandyopadhyay, S. Chatterjee, S. Dattagupta, A. Frydman, S. Sengupta, and P. A. Sreeram, Phys. Rev. B {\bf71}, 054401 (2005)
\bibitem{duttagupta2}M. Bandyopadhyay,  S. Dattagupta, Phys. Rev. B {\bf74}, 214410 (2006)
\bibitem{authiol}P. Crespo, R. Litrán, T. C. Rojas, M. Multigner, J. M. de la Fuente, J. C. Sánchez-López, M. A. García, A. Hernando, S. Penadés, and A. Fernández, Phys. Rev. Lett. {\bf 93}, 087204 (2004) 
\bibitem{hernando}A. Hernando, P. Crespo, and M. A. García, Phys. Rev. Letts, {\bf96}, 057206 (2006).
\bibitem{graphite}P. Esquinazi, D. Spemann, R. Höhne, A. Setzer, K.-H. Han, and T. Butz, Phys. Rev. Lett. {\bf91}, 227201 (2003), K. Kusakabe and M. Maruyama, Phys. Rev. B {\bf67}, 092406 (2003), P. O. Lehtinen, A. S. Foster, Yuchen Ma, A. V. Krasheninnikov, and R. M. Nieminen, Phys. Rev. Letters. {\bf93}, 187202 (2004) 
\bibitem{sangamapl} S. Banerjee, M. Mandal, N. Gayathri, and M. Sardar, Appl. Phys. Lett. {\bf91}, 182501 (2007)
\bibitem{oxides} M.Venkatesan, C. B. Fitzgerald and J.M.D. Coey, Nature {\bf430}, 630 (2004)
\bibitem{borides} L. S. Dorneles, M. Venkatesan, M. Moliner, J. G. Lunney, and J. M. D. Coey, Appl. Phys. Lett. {\bf85}, 6377 (2004)
\bibitem{simu}S.Banerjee, D.Bhattacharya, Computational Materials Science {\bf 44}, 41 (2008) \bibitem{srikanth1} S. Pal, M. Morales, P. Mukherjee and H. Srikanth, J. Appl. Phys. {\bf105}, 07B504 (2009).
\bibitem{du} J. Du, B. Zhang, R. K. Zheng, and X. X. Zhang, Phys. Rev. B {\bf75}, 014415 (2007)
\bibitem{bitoh} T. Bitoh, K. Ohba, M. Takamatsu, T. Shirane and S.
Chikazawa, Jour. Phys. Soc. Japan, {\bf64}, 1305(1995).
\bibitem{srikanth2} N. A. Frey, M. H. Phan, H. Srikanth, S. Srinath, C. Wang, and S. Sun, J. Appl. Phys., {\bf105}, 07B502 (2009)
\bibitem{wang}L. Wang, J.Luo, Q. Fan, M. Suzuki, I.S.Susuki, M.H.Engelhard,Y.Lin, N. Kim, J.Q.Wang, C-J. Zhong, J. Phys. Chem. B {\bf 109}, 21593 (2005)
\bibitem{caruntu} C. Caruntu, G. Caruntu and C. J. O'Conner,
Jour. Physics D {\bf40}, 5801(2007).
\bibitem{denardin} J. C. Denardin, A. L. Brandl, M. Knobel, P. Panissod, A. B. Pakhomov, H. Liu, and X. X. Zhang, Phys. Rev. B {\bf65}, 064422 (2002). 
\bibitem{djurberg}C. Dijurberg, P. Svedlindh, and P. Nordblad, M. F. Hansen, F. Bødker, and S. Mørup, Phys. Rev. Letts., {\bf79}, 5154 (1997).
\bibitem{wandersman} E. Wandersman, V. Dupuis, E. Dubois, R. Perzynski, S. Nakamae and E. Vincent, Euro. Phys. letts. {\bf84} 37011 (2008)
\bibitem{chaikin} P. M. Chaikin and T. C. Lubensky, Principle of
Condensed Matter Physics, CUP,1995.
\bibitem{mydo} J. A. Mydosh, Spin Glass: An experimental Introduction,
Taylor and Francis, London 1993.
\bibitem{blundell} R. E. Blundell, K. Humayn and A. J. Bray, Journal
of Physics A 25, L733 (1992).
\bibitem{bern} L. Bernardi and I. A. Campbell, Phys. Rev. B {\bf49}, 728(1994).
\bibitem{bianco}L. D. Bianco, D. Fiorani, A. M. Testa, E. Bonetti and L. Signorini, Phys. Rev. B, {\bf70}, 052401 (2004)
\bibitem{he} J.H. He, S.L. Yuan, Y.S. Yin, Z. M. Tian, P. Li, Y. Q. Wang, K. L. Liu and C. H. Wang, Jl. Appl. Phys. {\bf 103}, 023906 (2008)
\bibitem{kavich} D. W. kavich, J. H. Dickerson, S. V. Mahajan, S. A. Hasan and J.-H. Park, Phys. Rev B, {\bf78} 174414 (2008)
\bibitem{nogues} J. Nogues, J. Sort, V. Langlais, v. Skumryev, s. Surinach, J. S. Munoz, M. D. Baro, Phys. Rep. {\bf422} 65 (2005)
\bibitem{wang} H. Wang, T. Zhu, K. Zhao, W. N. Wang, C. S. Wang, Y. J. Wang, and W. S. Zhan, Phys. Rev. B {\bf70} 092409 (2004)
\bibitem{eftaxias} E. Eftaxias and K. N. Trohidou, Phys. Rev. B {\bf71} 134406 (2005)
\bibitem{solamon} D. Niebieskikwiat and M. B. Salamon, Phys. Rev. B {\bf72} 174422 (2005)
\bibitem{chen2} J. Wang, J. Sun, Q. Sun, Q. Chen, Mat. Res. Bull. {\bf38}, 1113 (2003)
\bibitem{surf}R. H. Kodama, A. E. Berkowitz, E. J. McNiff, Jr. and S. Foner, Phys. Rev. Letts. {\bf77}, 394 (1996), B. Martinez, X. Obrados, Ll. Balcells, A.Rpuanet and C. Monty, Phys. Rev. Letts. {\bf 80}, 181 (1998) 
\bibitem{hauser} J. J. Hauser, Phys. Rev. {\bf187}, 580(1969)
\bibitem{clog} A. M. Clogston, Phys. Rev. Letters, {\bf19}, 583(1967)
\bibitem{coau}F. Wilhelm, M. Angelakeris, N. Jaouen, P. Poulopoulos, E. Th. Papaioannou, Ch. Mueller, P. Fumagalli, A. Rogalev, and N. K. Flevaris, Phys. Rev. B {\bf69}, 220404(R) (2004)
\bibitem{Au} Y. Yamamoto,T. Miura, M. Suzuki, N. Kawamura, H. Miyagawa, T. Nakamura, K. Kobayashi, T. Teranishi, and H. Hori, Phys. Rev. Letters, {\bf93}, 116801.
\bibitem{carmeli} L. Carmeli, G. Leitus, R. Naaman, S. Reich, Z. Vager, J. Chem. Phys. {\bf118}, 10372(2003). 
\bibitem{lashell} S. Lashell, B. A. Mcdougall and E. Jensen, Phys. Rev. Letts, {\bf77}, 3419(1996)
\bibitem{koroteev} Yu. M. Koroteev, G. Bihlmayer, J. E. Gayone, E. V. Chulkov, S. Blügel, P. M. Echenique, and Ph. Hofmann, Phys. Rev. Letts. {\bf93}, 046403
\bibitem{sampedro} B. Sampedro, P. Crespo, A. Hernando, R. Litrán, J. C. Sánchez López, C. López Cartes, A. Fernandez, J. Ramírez, J. González Calbet, and M. Vallet, Phys. Rev. Letters, {\bf91}, 237203 (2003), T. Shinohara, T. Sato, and T. Taniyama, Phys. Rev. Letts, {\bf91}, 197201(2003).
\end{thebibliography}
\end{document}